\documentclass[prd,amsmath,amssymb,showpacs]{revtex4}
\usepackage{graphicx}
\usepackage{color}
\usepackage[colorlinks=true,linkcolor=red]{hyperref}
\usepackage{hyperref}

\begin{document}
\title{Fermion generations, masses and mixings in a 6D brane model}
\author{Silvestre Aguilar}
\email{sma34@csufresno.edu}
\author{Douglas Singleton}
\email{dougs@csufresno.edu} 
\affiliation{Physics Dept., CSU Fresno, Fresno, CA 93740-8031}

\date{\today}

\begin{abstract}
We study the motion of higher dimensional fermions in a non-singular 
6D brane background with an increasing warp factor. This background
acts as a potential well trapping fermions and fields of other spins
near a 3+1 dimensional brane. By adjusting the shape of this potential well
it is possible to obtain three normalizable zero mass modes
giving a possible higher dimensional solution to the fermion generation 
puzzle. The three different zero mass modes correspond to 
the different angular momentum eigenvalues for rotations
around the brane. This bulk angular momentum acts as the family or generation 
number. The three normalizable
zero modes have different profiles with respect to the bulk, thus
by coupling the higher dimensional fermion field to a higher dimensional
scalar field it is possible to generate both a realistic mass hierarchy and 
realistic mixings between the different families.
\end{abstract}


\maketitle


\section{Introduction}

In the standard model of particle physics \cite{SM} there are open questions
which have not yet found an answer. Chief among these is the fermion family or generation 
puzzle as to why the first generation of quarks and leptons (up quark, down quark,
electron and electron neutrino) are replicated in two families or generations
of increasing mass (the second generation consisting of charm quark, strange quark, muon and muon 
neutrino; the third generation consisting of top quark , bottom quark, tau and tau neutrino)
In addition to explaining why there are heavier copies of the first generation
of fermions one would like to explain the mass hierarchy of the generations and
the mixings between the generations characterized by the CKM (Cabbibo-Kobayashi-Maskawa) matrix. 
Several ideas have been suggested such as a horizontal family symmetry \cite{FS}.

Recently theories with extra dimensions have been used in a novel way to
try and explain some of the open questions in particle physics and cosmology.
In \cite {ADD} \cite{RS} \cite{Gog} the hierarchy problem (i.e.
why the gravitational interaction is many orders of magnitude weaker than the 
strong and electroweak interactions of particle physics) was addressed using large or 
infinite extra dimensions. Early versions of these extra dimensional models were
investigated by several researchers \cite{Ak} \cite{Ru} \cite{Vi} \cite{Gib}. 
In contrast to extra dimensions in the usual Kaluza-Klein picture, 
in the models with large or infinite extra dimensions
gravity acts in all the spacetime dimensions, while the other particles and fields
are confined, up to some energy scale, to a $3+1$ dimensional brane. 
These recent extra dimensional models have also been applied to answer other 
questions of particle physics and cosmology. Non-supersymmetric string models \cite{crem}
have been constructed which have been able to reproduce the standard model particles
from intersecting D5-branes. More phenomenological brane models \cite{Mass} \cite{neronov} \cite{Mi}
\cite{tait} have been constructed to explain the hierarchy of masses and/or the CKM elements of
the fermions. In \cite{DE} \cite{DM} brane world models were used to explain
dark energy and dark matter. General studies of higher dimensional cosmologies can be
found in \cite{Mel1} \cite{Mel2}.          

In this paper we attempt to give a toy model for the generation problem using a brane world
model in 6D. We obtain three 4D fermion families from the 
zero modes of a single 6D spinor field. A mass hierarchy and mixings between 
the three zero modes are obtained by introducing a Yukawa type interaction
between the 6D spinor field and a 6D scalar field. This gives a common origin 
(i.e. the higher dimensional Yukawa interaction) for both the mass
spectrum and the mixings of the fundamental fermions. 
 
\section{6D gravitational background}

In \cite{Mi} \cite{GoSi} \cite{GoSi1} \cite{Ma} a 6D brane world model was 
investigated which gave universal gravitational trapping of fields of
spins $0, \frac{1}{2}, 1, 2$, to the brane. The system considered was 6D gravity with
a cosmological constant and some matter field energy-momentum. The action for this
system was
\begin{equation} \label{action}
S = \int d^6x\sqrt{- ^6g}\left[\frac{M^4}{2}(^6R + 2 \Lambda) + ^6L \right] ~,
\end{equation}
where $\sqrt{-^6g}$ is the determinant, $M$ is the fundamental
scale, $^6R$ is the scalar curvature, $\Lambda$ is the
cosmological constant and $^6L$ is the Lagrangian of the matter
fields. All of these quantities are six dimensional.
The ansatz for the 6D metric was taken as
\begin{equation}\label{ansatzA}
ds^2  = \phi ^2 (r)\eta _{\alpha \beta } (x^\nu  )dx^\alpha
dx^\beta   - \lambda (r)(dr^2  + r^2d\theta ^2)~ ,
\end{equation}
where the Greek indices $\alpha, \beta,... = 0, 1, 2, 3$ refer to
4-dimensional coordinates. The metric of ordinary 4-space,
$\eta_{\alpha \beta }(x^\nu)$, has the signature $(+,-,-,-)$. The
functions $\phi (r)$ and $\lambda (r)$ depend only on the extra
radial coordinate, $r$, and thus are cylindrically symmetric in
the transverse polar coordinates ($0 \le r < \infty$, $0 \le
\theta < 2\pi$). The ansatz for the energy-momentum tensor of the matter 
fields was taken to have the form
\begin{equation} \label{source}
T_{\mu\nu} = - g_{\mu\nu} F(r), ~~~~~ T_{ij} = - g_{ij}K(r),
     ~~~~~ T_{i\mu} = 0 ~.
\end{equation} 
Other than satisfying energy-momentum conservation i.e.
\begin{equation}\label{energy-con}
\nabla^A T_{AB} = \frac{1}{\sqrt{-^6 g}}
\partial _A (\sqrt{-^6 g}T^{AB}) + \Gamma ^B _{CD} T^{CD} = 
K^{\prime} + 4\frac{\phi^\prime}{\phi} \left(K - F \right) = 0~,
\end{equation}
the energy-momentum tensor was unrestricted, although it was desirable for
it to satisfy physical requirements such as being everywhere finite, and 
being peaked  near the brane. In \eqref{energy-con} and in the rest of the paper
a prime indicates derivative with respect to $r$

In \cite{Mi} \cite{GoSi1} it was found that the above system had the following
non-singular solution
\begin{equation} \label{phi}
\phi (r) = \frac{c^b+ a r^b}{c^b+ r^b}~,
\end{equation}
where $a, b, c$ are constants and $a>1$. All other ansatz functions were given in
terms of $\phi (r)$.
\begin{equation}
\label{g} 
\lambda (r)= \frac{\rho^2 \phi ^{\prime}}{r} = \rho ^2 (a-1) b c^b  \frac{r^{b-2}}{(c^b+r^b)^2} ~,
\end{equation}
where $\rho$ is an integration constant with units of length, which was related to 
the constants $a$ and $b$ by
\begin{equation}
\label{ab}
\frac{\rho^2 \Lambda}{10 M^4} = \frac{b}{a-1} ~.
\end{equation}
This solution in terms of $\phi (r)$ and $\lambda (r)$ represents a non-singular, thick brane.
The brane thickness is proportional to $c$. 
Recently the matching conditions for a general thick brane of codimension 2 was given
\cite{jose}. This provides a general framework in which to study gravitational 
phenomena and particle trapping in codimension 2 braneworlds.
The source functions are also determined by $\phi (r)$.
\begin{equation} \label{FK}
F(r) = \frac{f_1}{2 \phi^2} +\frac{3 f_2}{4 \phi} ~, ~~~~~ K(r) =
\frac{f_1}{\phi^2} +\frac{f_2}{\phi} ~,
\end{equation}
where $f_1$ and $f_2$ are constants given by 
\begin{equation} \label{parameters}
 f_1 = -\frac{3\Lambda}{5} a~, ~~~~~ f_2=\frac{4 \Lambda}{5}(a+1)~,
\end{equation}
In \cite{Mi} and \cite{GoSi1} and it was shown that for $b=2$ this solution gave a
universal, gravitational trapping for fields with spins $0, \frac{1}{2} , 1, 2$ within the
brane width ($\approx c$) of $r=0$. 

For $b > 2$ \eqref{g} shows that the scale factor for
the extra dimensions, $\lambda (r) = 0$ at $r=0$, raising the possibility of having a
singularity on the brane and making the solutions with $b > 2$ unphysical. However,
by looking at invariants such as the Ricci scalar, $R$, one finds that they are 
non-singular at $r=0$. This indicates that the zero of $\lambda (r)$ at $r=0$ for $b >2$ solutions 
is not a physical singularity. The Ricci scalar for the above solution is
\begin{equation}
\label{ricci}
R = \frac{2b (-5 c^{2 b} + 4 a c^{2 b} + 10 c^b r^b - 
22 a c^b r^b + 10 a^2 c^b r^b + 4 a r^{2 b} - 5a^2 r^{2 b})}{(a-1)\rho^2 (c^b + a r^b)^2} ~.
\end{equation}
It is easy to see that this is finite at $r=0$. Other invariants such as the fully contracted
Riemann tensor, $R_{AB} ~^{CD} R_{CD} ~^{AB}$, or the square of the Ricci tensor, $R_{AB} R^{AB}$
also turn out to be finite at $r=0$. Thus we take the zero in the scale factor $\lambda (r)$ to
be a coordinate rather than physical singularity. Note that $\lambda (r)$ goes to zero both at
$r=0$ and $r= \infty$ so that the metric \eqref{ansatzA} essentially becomes 4D at these
locations. Thus the solution given in \eqref{phi} is like the 2 brane model of \cite{RS}
where two 4D branes sandwich the higher dimensional, bulk spacetime.  

The weak point in the above is that the source ansatz functions have no clear physical 
interpreation. It would be desirable to show that an energy-mometum tensor of the
general form given by $F(r), K(r)$ could arise from some realistic source 
such as a scalar field. Because $\phi (r)$ increases as $r \rightarrow \infty$
\eqref{FK} \eqref{parameters} indicate that $F(r), K(r)$ have their maximum near $r=0$
and decrease as $\rightarrow \infty$. This behavior is similar to soliton solutions of classical
field theory. Thus $F(r), K(r)$ might be considered as modeling some solitonic field
configuration which forms the brane. As specific examples references \cite{Br} \cite{Br1} 
investigate solitonic scalar field configurations which form branes. Another possible physical
interpretation of the source ansatz functions can be given by transforming the metric given by
\eqref{ansatzA} \eqref{phi} \eqref{g} via the transformation $r=c \tan^{2/b} (\beta /2)$
and setting $c^2 /4 =10 M^4 / \Lambda =\rho ^2 (a-1)/ b$. With this the metric takes
the form
\begin{equation}
\label{cylinder}
ds^2  = \phi ^2 (\beta)\eta _{\alpha \beta } (x^\nu  )dx^\alpha
dx^\beta   - \frac{c^2}{4} \left( d\beta^2  + \frac{b^4}{4} \sin ^2 \beta d\theta ^2 \right)~ ,
\end{equation}
where $\phi (\beta) = \frac{1}{2} ((a+1) +(1-a) \cos \beta)$. When $b=2$ one can see
(after renaming the angles in a standard way as $\beta \rightarrow \theta$ and
$\theta \rightarrow \varphi$) that the geometry of the extra dimensions is that of a sphere.
However when $b>2$ one has zeroes in the 2D scale function, $\lambda (r)$, 
at $r=0$ and $r=\infty$ and a conical deficit angle of $\delta = 2 \pi - b \pi$.
This is essentially the ``football''-shaped geometry for the extra dimensions 
considered in \cite{carroll} (see in particular equation (24) of \cite {carroll}).
The 6D solution in \cite{carroll} involved realistic sources: an electromagnetic field in the 
form of a magnetic flux and a bulk cosmological constant. Thus the source ansatz functions,
$F(r) , K(r)$ may be acting as effective magnetic flux plus bulk cosmological constant.  

\section{Three families from three zero modes}

We now study the motion of a 6D fermion field in the gravitational background
given by \eqref{ansatzA}. We make the identification that different fermion zero-mass modes
(i.e. solutions for which the 4D part of the fermion wavefunction satisfies
$\gamma _\mu \partial ^\mu \psi (x _\nu ) = 0$) correspond to different families.
Thus we want to see if it is possible to obtain three zero modes. The picture that 
we present here is a toy model in the sense that the effective 4D fermion fields do not have
the full $SU(3) \times SU(2) \times U(1)$ charges of the real Standard
Model fields. As will be shown below the fermions considered here carry only a $U(1)$ 
charge associated with the rotational symmetry around the brane i.e. the symmetry
associated with the extra dimensional variable $\theta$. This is the same as the first 
example in \cite{neronov}, with the $U(1)$ quantum number being associated with
family number. Reference \cite{neronov} gives other more
complex and realistic examples where the fermions carry $U(1) \times U(1)$ or
$SU(2) \times U(1)$ charges. In the $U(1) \times U(1)$ example one of the
$U(1)$'s is associated with a horizontal, family symmetry \cite{FS} which distinguishes between
fermions of different generations, while the other $U(1)$ is then an ordinary gauge symmetry.
In the present case the single $U(1)$ is associated with the family symmetry.  
Other authors \cite{hung} also have studied 6D brane models with more realistic standard model charges.

One can gauge \cite{neronov} any of the above examples using the Kaluza-Klein approach. 
For example, with the simple $U(1)$ model considered here the metric in \eqref{ansatzA} has 
a Killing vector, $\partial _\theta$, which via the standard Kaluza-Klein mechanism implies an
associated gauge boson in the effective 4D theory. The gauge field arises from the off-diagonal
components of the higher dimensional metric as
\begin{equation}
\label{kk-gauge}
ds^2  = \phi ^2 (r)\eta _{\alpha \beta } (x^\nu  )dx^\alpha
dx^\beta   - \lambda (r)(dr^2  + (r d\theta + A_\mu dx^\mu)^2 )~ .
\end{equation} 
This $A_\mu$ is analogous to the horizontal gauge boson of the family symmetry models
\cite{FS}. In \cite{neronov2} it was shown that the zero mode of the
gauge field $A_\mu$ was localized to the brane. Since horizontal gauge
bosons have an experimentally fixed lower mass limit a more realistic model would
need to have some symmetry breaking,  Higgs mechanism in order to give $A_\mu$ an acceptably
large mass. 

From the previous section we found that in 6D one has the freedom to
let the exponent $b$ take values other than $b=2$. The motion of fermions in 
the 6D brane solution of \eqref{ansatzA} for the $b=2$ case was
studied in \cite{Ma}. It was found that in this case only one
zero mode occurred, and thus only one family. Therefore we want to 
consider the $b>2$ case and show that for a certain range of $b$ it is possible
to get three zero modes. The constant $b$ controls the steepness of the scale functions
$\phi (r)$ and $\lambda (r)$. Therefore it is reasonable that larger $b$ should give a stronger
confinement of the fermions to the vicinity of the brane at $r=0$, and thus to a larger number of
zero modes.

The 6D action and resulting equations of motion for a spinor field are
\begin{equation}
\label{fermion}
S_\Psi = \int d^6 x \sqrt{-^6g}\bar{\Psi} i \Gamma^A D_A \Psi
~~, \qquad \Gamma ^A D_A \Psi = \Gamma ^{\mu} D_{\mu} \Psi + \Gamma ^r D_r \Psi +
\Gamma ^{\theta} D_{\theta} \Psi = 0  ~.
\end{equation}
In the above $\Gamma ^A = h^A _{\bar B} \gamma ^B$ are the 6D curved spacetime
gamma matrices, and $h^A _{\bar B}$ are the sechsbiens defined via 
$g_{AB} = h ^{\bar A} _A h^{\bar B} _B \eta _{\bar A \bar B}$. In order to
evaluate the 6D Dirac equation in \eqref{fermion} we need to calculated the
spin connections
\begin{equation} \label{spin1}
\omega^{\bar{M}\bar{N}}_M = \frac{1}{2} h^{N\bar{M}} (\partial_M
h^{\bar{N}}_N - \partial_N h^{\bar{N}}_M) - \frac{1}{2}
h^{N\bar{N}}(\partial_M h^{\bar{M}}_N - \partial_N h^{\bar{M}}_M)
- \frac{1}{2} h^{P\bar{M}}h^{Q\bar{N}}(\partial_P h_{Q\bar{R}} -
\partial_Q h_{P\bar{R}})h^{\bar{R}}_M ~.
\end{equation}
The non-zero spin connections are
\begin{equation} \label{spin2}
\omega^{\bar{r}\bar{\nu}}_\mu = \delta ^{\bar{\nu}} _{\mu} \frac{\sqrt{r \phi '}}{\rho} ~ , ~~~~~
\omega^{\bar{r}\bar{\theta}}_\theta = \sqrt{\frac{r}{\phi '}} \partial _r 
\left( \sqrt{r \phi' } \right) ~.
\end{equation}
With these one can explicitly calculate the various covariant derivatives in \eqref{fermion}
\begin{equation} \label{covariant}
D_\mu\Psi = \left( \partial_\mu +  \frac{1}{2} \omega^{\bar{r}\bar{\nu}}_\mu 
\gamma _r \gamma _\nu \right) \Psi ~ , ~~~~~ D_r\Psi = \partial_r \Psi ~ , ~~~~~
D_\theta \Psi = \left(\partial_\theta + \frac{1}{2} \omega^{\bar{r}\bar{\theta}}_\theta 
\gamma _r \gamma _\theta \right) \Psi ~.
\end{equation}
We now assume that the 6D fermion spinor can be decomposed as $\Psi (x^A) = \psi (x_\mu ) \otimes
\zeta (r, \theta)$ into 4D and 2D parts. We are interested in the zero-mass modes so the 4D fermion
part is taken to satisfy $\gamma _\mu \partial ^\mu \psi (x _\nu ) = 0$. The 2D spinor can be expanded
as 
\begin{equation}
\label{2Dspinor}
\zeta  (r, \theta ) =  
\left(
\begin{array}
[c]{c}
f_l (r) \\
g_l (r)
\end{array}
\right)
e^{i l \theta} 
\end{equation}
We take the gamma matrices of the extra space as in \cite{neronov} \cite{Mi}
\begin{equation}
\label{2dgamma}
\gamma ^r  =\left(
\begin{array}
[c]{cc}%
0 & 1\\
1 & 0 
\end{array}
\right) \qquad
\gamma ^\theta  =\left(
\begin{array}
[c]{cc}%
0 & -i\\
i & 0 
\end{array}
\right) .
\end{equation}
Combining equations \eqref{fermion} - \eqref{2dgamma} we arrive at the following 
equations for $f_l (r)$ and $g_l (r)$
\begin{equation}
\label{fg-fermion}
\left[ \partial_r + 2 \frac{\phi '}{ \phi} + \frac{1}{2} 
\frac{\partial_r \left( \sqrt{r \phi ' } \right)}{\sqrt{r \phi'}}   
+ \frac{l}{r} \right] g_l(r) = 0  ~, \qquad
\left[ \partial_r + 2 \frac{\phi '}{ \phi} + \frac{1}{2} 
\frac{\partial_r \left( \sqrt{r \phi ' } \right)}{\sqrt{r \phi'}}   
- \frac{l}{r} \right] f_l(r) = 0 ~.
\end{equation}
The solutions for $f_l (r)$ and $g_l (r)$ are
\begin{equation}
\label{fg-solution}
f_l (r)  =  a_l \phi (r) ^{-2} (r \phi ' (r) )^{-\frac{1}{4}} r^l ~ , ~~~~
g_l (r)  =  b_l \phi (r) ^{-2} (r \phi ' (r) )^{-\frac{1}{4}} r^{-l} ~.
\end{equation}
Because of the $l$ dependence in both $f_l (r)$ and $g_l (r)$, different
$l$ values give fermion fields with different profiles in the bulk.
This will result in different masses and mixings for different $l$ values
when the fermion field is coupled to a scalar field in the next section.
One criteria for the trapping of the fermion field is that it should be
normalizable with respect to the extra dimensions $r, \theta$
\begin{equation}
\label{norm}
1 = \int \sqrt{-^6 g}  {\bar \zeta (r, \theta)} \zeta (r, \theta) dr d \theta = \int _0 ^{2 \pi} d \theta
\int _0 ^\infty dr \phi ^4 \rho ^2 \phi '\frac{\phi ^{-4}}{\sqrt{r \phi '}}
(a_l^2 r^{2l} + b_l ^2 r^{-2l}) =
2 \pi \rho ^2 \int _0 ^\infty dr \sqrt{\frac{\phi '}{r}} (a_l ^2 r^{2l} + b_l ^2 r^{-2l})
\end{equation}
From \eqref{g} one finds $\sqrt{\phi ' /r} = (a-1)^{\frac{1}{2}} b^{\frac{1}{2}} c^{\frac{b}{2}} 
r^{\frac{b}{2} -1} (c^b + r^b)^{-1}$.
Thus in order for \eqref{norm} to be normalizable and for the particular fermion $l$-mode
to be trapped, we want the integral
\begin{equation}
\label{norm2}
2 \pi \rho ^2 \sqrt{(a-1) b c^b} 
\int _0 ^\infty \frac{r^{\pm 2 l + \frac{b}{2} -1}}{c^b + r^b} dr ~,
\end{equation}
to be finite. If \eqref{norm2} diverges the particular $l$-mode will not
be trapped. This requirement that \eqref{norm2} be finite leads to restrictions
on $b$ for particular values of $l$. Evaluating the integral \eqref{norm2}
gives
\begin{equation}
\label{norm3}
2 \pi^2 \rho ^2 c^{\pm 2l} \sec\left(\frac{2l\pi}{b} \right) \sqrt{\frac{a-1}{b}} 
\qquad \text{if} \qquad b > 4|l|,
\end{equation}
and \eqref{norm2} diverges if $b \le 4 |l|$. 
Thus in order to have three normalizable $l$-modes we require that
$4 < b \le 8$. Under these conditions the $l=0$, and $|l|=1$ modes are normalized
and trapped, while $|l| \ge 2$ modes are not. Since the integrand in \eqref{norm2} is positive definite
and only has possible divergences at $r=0$ and $r=\infty$ one can come to this conclusion 
by investigating the $r \rightarrow 0$ and $r \rightarrow \infty$ behavior of this 
integrand. For $|l| =1$ one finds that for $l=+1$ the integrand behaves as 
$^{lim} _{r \rightarrow 0} \simeq r^{1 + \frac{b}{2}}$ , 
$^{lim} _{r \rightarrow \infty} \simeq r^{1 - \frac{b}{2}}$ ; for $l=-1$, it 
behaves as $^{lim} _{r \rightarrow 0} \simeq r^{-3 + \frac{b}{2}}$,
$^{lim} _{r \rightarrow \infty} \simeq r^{-3 - \frac{b}{2}}$. The $r \rightarrow 0$
limit of $l=+1$ and $r \rightarrow \infty$ limit of $l=-1$ give convergent results. 
On the other hand the $r \rightarrow 0$ limit of $l=-1$ and $r \rightarrow \infty$ limit of $l=+1$ 
give convergent results only if $b>4$. One can see the for $b>4$ the $l=0$ mode
is normalized. For $|l|=2$ one finds that for $l=+2$ the integrand behaves as 
$^{lim} _{r \rightarrow 0} \simeq r^{3 + \frac{b}{2}}$ , 
$^{lim} _{r \rightarrow \infty} \simeq r^{3 - \frac{b}{2}}$ ; for $l=-2$, it 
behaves as $^{lim} _{r \rightarrow 0} \simeq r^{-5 + \frac{b}{2}}$,
$^{lim} _{r \rightarrow \infty} \simeq r^{-5 - \frac{b}{2}}$.
The $l=+2$ integral diverges at $r \rightarrow \infty$ and the $l=-2$ diverges at
$r \rightarrow 0$ if $b \le 8$. This analysis again shows that one has three normalizable modes
(i.e. three fermion families) when the $4 < b \le 8$.

One could also consider using the criteria for trapping that the fermion action be finite when integrated over
the extra dimensions. 
\begin{equation}
\label{f-action}
S_\Psi = \int d^6 x \sqrt{-^6g}\bar{\Psi} i \Gamma^A D_A \Psi = 
2 \pi \rho ^2 \int _0 ^\infty dr \frac{1}{\phi} \sqrt{\frac{\phi '}{r}} (a_l ^2 r^{2l} + b_l ^2 r^{-2l}) ~.
\int d^4 x \sqrt{-\eta} {\bar \psi} i \gamma ^\nu \partial _\nu \psi
\end{equation}
The fermions are trapped if the integral over $r$ in the last expression is convergent. This integral
is almost the same as the last integral in \eqref{norm}. It differs only by a factor of $1 / \phi$ which
comes from the sechsbien that modifies the gamma matrices, $\gamma ^\nu$. The explicit expression 
for $\phi ^{-1} \sqrt{\phi ' /r}$ can be read off from \eqref{g} and \eqref{phi}. From
this one finds that $\phi ^{-1} \sqrt{\phi ' /r} \propto r^{\frac{b}{2} -1} (c^b + a r^b)^{-1}$.
The only change with respect to the normalization condition \eqref{norm} is that $r^b \rightarrow a r^b$ 
in the denominator. Thus the integral of the  action over the extra coordinates will have 
the same convergence properties as the normalization condition \eqref{norm} , thus giving the same conclusion 
that three zero-mass modes will be trapped if $4<  b \le 8$.

In \cite{Ma} only the $b=2$ case in \eqref{phi} was considered and only one zero-mass mode occurred.
Thus the existence of three zero modes is the result of allowing the exponent in \eqref{phi}
to take values $b>2$.  In \cite{Si} it was shown that the solution of \eqref{phi} \eqref{g} \eqref{FK} 
could be generalized to spacetimes of dimension greater than 6D. For these greater than 6D spacetimes the 
exponent, $b$, in \eqref{phi} was not free, but fixed to $b=2$. This would seem to imply that
only in 6D can one have more than one fermion generation for the background solution given by
\eqref{phi} \eqref{g} \eqref{FK}. However in the case where spacetime greater than 6D 
one could consider taking the higher generations as non-zero mass modes. Also one could try to 
generalize the other 6D brane solution given in \cite{Ma} to spacetime dimensions greater than 6D.

In discussing the masses and mixing between the different families (i.e. different $l$) we will
need the normalization relationship between $a_l$ and $b_l$. From \eqref{norm} \eqref{norm2} and \eqref{norm3}
we find 
\begin{equation}
\label{norm4}
a_l ^2 c^{2l} + b_l^2 c^{-2l} =  \frac{\cos \left(\frac{2 l \pi}{b} \right)}{2 \pi ^2 \rho ^2} \sqrt{\frac{b}{a-1}} ~.
\end{equation}
This condition allows us to write $b_l$ in terms of $a_l$ or visa versa. 

\section{Mixings and Masses}

By adjusting the exponent $b$ in our gravitational background solution we have three zero mass
modes which can be taken as a toy model for three generations of fermions. 
There are two problems: first there is no mixing between the different 
generations due to the orthogonality of the angular parts of the
higher dimensional wave functions. Overlap integrals like 
$\int_0 ^\infty \int _0 ^{2 \pi} {\bar \zeta _l} \zeta _m dr d \theta$, which characterize the mixing
between the different states, vanish since $\int _0 ^{2 \pi} e^{-il\theta} e^{im\theta} d \theta =0$
if $l \ne m$. Second, all the states are massless, whereas the fermions
of the real world have masses that increase with each succeeding family. 
Following \cite{neronov} we address both of these issues
by introducing a coupling between the 6D fermions and a 6D scalar field of the form
$H_p (x ^A ) {\bar {\Psi}} _l (x^B) \Psi _{l'} (x^C )$. This adds to the action a scalar-fermion
interaction of the form
\begin{equation}
\label{FFS}
S_{sf} = f \int d^4 x dr d\theta \sqrt{- ^6 g} H_p \bar {\Psi} _l \Psi _{l'} ~,
\end{equation}
$f$ is a constant which gives the scalar-fermion coupling strength.

We now take the scalar field to be of the form
\begin{equation}
\label{higgs}
H_p (x ^A) = H_p (r) e^{ip \theta} 
\end{equation}
i.e. only depending on the bulk coordinates $r , \theta$, but not on the brane
coordinates $x ^\mu$. In \cite{neronov} the same form as in \eqref{higgs} was
taken for the scalar field, but certain simplifying assumptions were made about the form
of $H(r)$ -- it was assumed to be either a constant or a delta-function.
In \cite{tait} other forms for the scalar field profile were used. 
In the following we will determine the form of $H(r)$ by studying the field
equations for a scalar field in the background provided by \eqref{phi} \eqref{g}. 
Note that in the form \eqref{higgs} the scalar field is only dynamical with respect to the 
extra dimensions, $r, \theta$, but not with respect to the brane spacetime dimensions,
$x^\mu$. Thus one has a scalar field mechanism for fermion mass generation without a
dynamical 4D scalar particle.

Substituting \eqref{higgs} into \eqref{FFS} we find  
\begin{equation}
\label{FFS2}
S_{sf} = U_{l l'} \int d^4 x \bar{ \psi} _l (x ^\mu ) \psi _{l'} ( x ^\mu) \qquad \text{where} \qquad
U_{ll'} = f \int dr d \theta \sqrt{- ^6 g} H_p (r) e^{i (p-l+l') \theta} \bar{ \zeta _l }(r) \zeta _{l'}
(r ) ~.
\end{equation}
$U_{ll'}$ will be non-zero when $p-l+l' = 0$. When $l=l'$ this gives a mass term and when 
$l \ne l'$ this gives a mixing term between the $l$ and $l'$ modes.  

To get explicit results for $U_{ll'}$ one needs an explicit form for $H_p (r)$. This is done
by solving the field equations for a test scalar field in the background given by $\phi (r)$ and
$\lambda (r)$ for the different $p-$modes. The equation for the scalar field in the background given
by \eqref{phi} \eqref{g} is
\begin{equation}
\label{eqn-sca}
\frac{1}{\sqrt{- ^6 g}} \partial _A \left(\sqrt{- ^6 g} g^{AB} \partial _B H_p (x^A ) \right) = 0 ~.
\end{equation}
Inserting \eqref{higgs} into \eqref{eqn-sca} we get
\begin{equation}
\label{eqn-sca2}
H_ p '' (r) + \left( \frac{1}{r} + \frac{4 \phi ' (r)}{\phi (r)} \right) H_p ' (r) -
\frac{p^2}{r^2} H_p (r) = 0 ~.
\end{equation}
From \eqref{phi} we find that
\begin{equation}
\label{4pp}
\frac{4 \phi ' (r)}{\phi (r)}= 
\frac{4 (a-1) \frac{b}{c} \left( \frac{r}{c}\right)^{b-1}}{\left(1 + \left( \frac{r}{c}\right)^b \right)
\left(1 + a\left( \frac{r}{c}\right)^b \right)} ~.
\end{equation}
Inserting this expression back into \eqref{eqn-sca2} we were not able to find a closed 
form solution, but were only able to solve it numerically. 
From the form of $4 \phi ' (r) / \phi (r)$ in \eqref{4pp} one can
see that when $ 4 < b \le 8$ that the $-p^2/r^2$ and $1/r$ terms dominate in the limits
$r \rightarrow 0$ and $r \rightarrow \infty$. If one drops the $4 \phi ' (r) / \phi (r)$ term
then one finds that the asymptotic ($r \rightarrow 0$ and $r \rightarrow \infty$) solutions to
\eqref{eqn-sca2} are
\begin{eqnarray}
\label{scalar-soln}
H_p (r) = A_{+ p} r^{ |p|} ~~~ \text{or} ~~~ A_{-p} r^{-|p|} ~~~ \text{for} ~~ p \ne 0  \\
H_0 (r) = A_0 ~~ \text{or} ~~ B_0 \ln (r) ~~~ \text{for} ~~ p=0 ~,
\end{eqnarray}
where $A_{\pm p}$, $A_0$, and $B_0$ constants. These asymptotic solutions gave a 
fair representation to the numerical solution even for intermediate values of $r$.
The $p=0$ solutions can be written in combined
form as $H_0 (r) = B_0 \ln (r/c_0)$ where $A_0 = - B_0 \ln (c_0)$. This form will be used
the next subsection to give masses to the three zero modes. The singularities in $H_0 (r)$, at
$r=0$ and $r= \infty$, are not a problem since the combination of the fermion ``wave function"
and the metric ansatz functions go to zero fast enough at $r=0$ and $r= \infty$ to negate
these singularities in the Yukawa coupling integral \eqref{FFS2}.

\subsection{Masses}

From \eqref{FFS2} one can sees that the mass terms are those for which $l=l'$ and thus 
we want to consider couplings to the $p=0$ scalar mode in \eqref{scalar-soln}. We will use
the combined form of the two solutions namely $H_0 (r) = B_0 \ln (r/ c_0)$. With this 
\eqref{FFS2} becomes
\begin{eqnarray}
\label{mass3}
m_l = U_{ll} &=&  f \int dr d \theta \sqrt{- ^6 g} H_0 (r) \bar{ \zeta _l }(r) 
\zeta _{l} (r )
= 2 f B_0 \pi \rho ^2 \sqrt{(a-1) b c^b} \int _0 ^\infty 
\frac{ \ln \left( \frac{r}{c_0} \right)r^{b/2 -1}}{c^b+r^b} \left[a_l^2r^{2l} + b_l^2 r^{-2l} \right] dr  \nonumber \\
& = & f B_0 \left( \ln \left( \frac{c}{c_0} \right) +  \frac{\pi K_l}{b} \tan \left( \frac{2 l \pi}{b} \right) \right)
\qquad \text{where} \qquad 
K_l = \frac{4 a_l^2 c^{2l} \pi ^2 \rho ^2}{\cos \left(\frac{2 \pi l }{b} \right)} \sqrt{\frac{a-1}{b}} -1 ~.
\end{eqnarray}
In arriving at the final line in \eqref{mass3} where have used \eqref{norm4} to replace $b_l$
in terms of $a_l$. Looking at only the $\tan (2 l \pi / b)$ term and taking $f B_0 K_l \pi / b >0$ gives a 
hierarchy of masses of $m_{-1} < m_0 < m_{+1}$. However $m_{-1} <0$ and $m_0 = 0$ which is phenomenologically 
wrong. Taking $\ln (c/ c_0)$ as positive (i.e. $c>c_0$) and sufficently large can shift 
the mass spectrum so that all masses are positive and while still
maintaining the hierarchy $m_{-1} < m_0 < m_{+1}$. Because 
our fermions only carry a $U(1)$ charge the above hierarchy is a toy model. 
Here, as an example, we take the three fermions as the ``down" sector of quark generations where $l=-1$
is the down quark, $l=0$ is the strange quark, and $l=+1$ is the bottom quark. Taking
the masses of the down, strange and bottom quarks as $m_{-1}= m_d =4$ MeV , $m_0 = m_s = 100$ MeV and
$m_{+1} = m_b = 4400$ MeV we find from \eqref{mass3} 
\begin{equation}
\label{mass4}
\frac{m_d}{m_s} = \frac{m_{-1}}{m_0} = 0.04 = \left( \frac{-K _{-1} \pi \tan ( 2 \pi / b )}{b \ln ( c/ c_0 )}
+1 \right)  ~ , ~~~~ \frac{m_b}{m_s} = \frac{m_{+1}}{m_0} = 44.00 = \left( \frac{ K_{+1} \pi \tan ( 2 \pi /b )}{b \ln ( c/ c_0 )}
+1 \right) ~.
\end{equation}
Solving these equations for $a_1$ and $a_{-1}$ (which are embedded in the definition of
$K_{+1}$ and $K_{-1}$) gives
\begin{equation}
\label{a1}
a_1 = \frac{D}{c^2} \sqrt{(1 + 43 x)\cos \left(\frac{2 \pi}{b} \right)}  ~~~, ~~~
a_{-1} = D \sqrt{(1 + 0.96 x)\cos \left(\frac{2 \pi}{b} \right)} ~,
\end{equation}
where $x=\frac{b \ln (c/c_0)}{\pi \tan (2 \pi /b)}$ and $D ^2 = \frac{c^2}{4 \pi^2 \rho ^2} \sqrt{\frac{b}{a-1}}$. If
$a_1$ and $a_{-1}$ are chosen as in \eqref{a1} then the mass ratios in \eqref{mass4} are
obtained. 

\subsection{Mixings}

A similar analysis can be carried out with the mixings between the different ``families" 
characterized by different $l$ number. The mixings are delineated by $U_{0,1} , U_{1,0} , 
U_{1, -1} , U_{-1, 1} , U_{0, -1}$ and $U_{-1, 0}$. In the case of mixings the scalar field
must have a non-zero angular eigenvalue i.e. $H_p (r , \theta) = H_p (r) e^{i p \theta}$
with $p \ne 0$) which satisfies $p-l+l' = 0$. Thus for $U_{-1, 0}$ and $U_{0,1}$ one needs $p=-1$;
for $U_{1, 0}$ and $U_{0, -1}$ one needs $p=1$; for $U_{1, -1}$ one needs $p=2$; for $U_{-1 , 1}$
one needs $p=-2$. We will require the following relationship between the mixings: $U_{0,1} = U_{1,0}$, 
$U_{1, -1} = U_{-1, 1}$ and $U_{0, -1} = U_{-1, 0}$. This in turn implies that $H_p (r)$ should
depend only on $|p|$ ({\it e.g.} $H_1 (r) = H_{-1} (r)$). Looking at the first line
in \eqref{scalar-soln} this means that we can take either the first solution or the
second but not the sum in general unless $A_{+1} = A_{-1}$. In what follows we will take
$H_p = A_{+p} r^{|p|}$. The conclusions are not qualitatively different if we make
the other choice $H_p = A_{-p} r^{-|p|}$. With this we find
\begin{equation}
\label{ckm}
U_{ll'} = f \int dr d \theta \sqrt{- ^6 g} H_p (r) \bar{ \zeta _l }(r) 
\zeta _{l'} (r ) = 2 \pi \rho ^2 f \sqrt{(a-1) b c^b} \int _0 ^{\infty} H_p (r)
\frac{r^{\frac{b}{2} -1}}{c^b+r^b} \left[a_l ^* a_{l'} r^{l+l'} + b_l ^* b_{l'} r^{-l-l'} \right] ~,
\end{equation}
where in \eqref{ckm} we have carried out the $d \theta$ integration, and the condition
$p-l+l'$ holds. Inserting $H_p (r) = A_{+p} r^{|p|}$ in \eqref{ckm} and assuming all $a_l$ and
$b_l$ are purely real we get
\begin{eqnarray}
\label{ckm2}
U_{ll'} &=& 2 \pi \rho ^2 f \sqrt{(a-1) b c^b} A_{+p} \int _0 ^{\infty} \frac{r^{\frac{b}{2}+p-1}}{c^b +r^b}
\left[ a_l a_{l'} r^{l+l'} + b_l b_{l'} r^{-l-l'} \right] \nonumber \\
&=& 2 \pi^2 \rho ^2 f \sqrt{\frac{a-1}{b}} A_{+p} c^{|p|} \left[
a_l a_{l'} c^{l+l'} \sec \left( \frac{(l+l'+|p|) \pi}{b} \right) +
b_l b_{l'} c^{-l-l'} \sec \left( \frac{(l+l'-|p|) \pi}{b} \right) \right]~.
\end{eqnarray}
Now the four cases ($l=0, ~ l'=-1$), ($l=-1,~ l'=0$), ($l=0, ~ l'=1$) and ($l=1,~ l'=0$) involve $|p|=1$
and from \eqref{ckm2} yield 
\begin{eqnarray}
\label{ckm3}
U_{0,-1} &=& U_{-1,0} =  \frac{f D A_{+1} c^2}{2} \left( \frac{a_0 a_{-1}}{c^2} + b_0 b_{-1} \sec \left( \frac{2 \pi}{b} \right)
\right)  \\
\label{ckm3a}
U_{0,1} &=& U_{1,0} = \frac{f D A_{+1} c^2}{2} \left( \sec \left( \frac{2 \pi}{b} \right) a_0 a_{-1} +  \frac{b_0 b_{-1}}{c^2}
\right)
\end{eqnarray}
For the two cases $l=1,~ l'=-1$, $l=-1,~ l'=1$ one has $|p|=2$
\begin{equation}
\label{ckm4}
U_{1,-1} = U_{-1,1} =  \frac{f D A_{+2} c^2}{2} \left(a_1 a_{-1}  + b_1 b_{-1} \right) \sec \left( \frac{2 \pi}{b} \right)~.
\end{equation}
where as in the previous subsection $D ^2 = \frac{c^2}{4 \pi^2 \rho ^2} \sqrt{\frac{b}{a-1}}$. 
We now show that parameters (i.e. $a, b, c, a_{0}$) can be chosen so the
ratios of the above mixings match the ratios of the CKM mixing matrix elements. If this 
can be done then taking $a_{\pm 1}$ as in \eqref{a1} will yield the correct mass
ratios. From \cite{pdb} we take $U_{0, -1} = V_{us} \approx 0.224$, 
$U_{0, 1} = V_{cb} \approx 0.040$ and $U_{-1, 1} = V_{ub} \approx 0.0036$. Again note that since
in our model we only have one flavor in each family (here taken as the ``down''
flavor or sector) these associations between $U_{i , j}$ and $V_{ij}$ are to be taken as
representing generic inter-family mixing. Now combining \eqref{ckm3} \eqref{ckm3a} 
\eqref{ckm4} gives
\begin{eqnarray}
\label{ckmratio}
\frac{U_{0, -1}}{U_{0,1}} = \frac{a_0 a_{-1} + c^2 \sec \left( \frac{2 \pi}{b} \right) b_0 b_{-1}}
{b_0 b_1 + c^2 \sec \left( \frac{2 \pi}{b} \right) a_0 a_1} \approx 5.6 \\
\label{ckmratioa}
\frac{U_{0, -1}}{U_{-1,1}} = \frac{ A_{+1} \left( \frac{a_0 a_{-1}}{c^2} + b_0 b_{-1} \sec \left( \frac{2 \pi}{b} \right)
\right)}{A_{+2}\left(a_1 a_{-1}  + b_1 b_{-1} \right) \sec \left( \frac{2 \pi}{b} \right)} 
\approx 60.5 ~.
\end{eqnarray}  
To determine the last ratio one needs to determine the normalization constants, $A_{+1}$ and $A_{+2}$ of
the $p=1$ and $p=2$ scalar field modes. For this one needs to explicitly evaluate the 
integral $1= A_p ^2 \int \sqrt{- ^6g} H_p ^2 dr d \theta$ for $p=1$ and $p=2$, with $H_p (r)$ given
by \eqref{scalar-soln}. Explicitly the ratio $A_{+1} / A_{+2}$ is
\begin{equation}
\label{a+1a+2}
\frac{A_{+1}}{A_{+2}} = c \sqrt{ \frac{\int _0 ^\infty \frac{(1+ay^b)^4}{(1+y^b)^6} y^{b+3}dy}
{\int _0 ^\infty \frac{(1+ay^b)^4}{(1+y^b)^6} y^{b+1}dy}} ~.
\end{equation}
The two mass ratios have already been fixed by choosing $a_{\pm 1}$ (this also
fixes $b_{\pm 1}$ because of \eqref{norm4}) as in \eqref{a1}. Next the mixing 
ratio, $\frac{U_{0, -1}}{U_{0,1}}$, in \eqref{ckmratio} can be fixed by choosing $a_0$ 
(this also fixes $b_0$ because of \eqref{norm4}). In particular let us parameterize
$a_0$ as $a_0 =\frac{D}{c} \eta$ so that $b_0 = \frac{D}{c} \sqrt{2 - \eta ^2}$ where 
$\eta$ is arbitrary, and $D$ and $c$ are previously defined. In this way 
one can see that \eqref{ckmratio} and also \eqref{a1} are independent of $c$. Finally one
can fix \eqref{ckmratioa} by choosing $a$ in \eqref{a+1a+2} to give the
ratio $A_{+1}/A_{+2}$. As an example
choose $a=7.5$, $b=4.041$, $c=1.0$, $D=1.0$ (this can be done by adjusting $\rho$), and $x=0.001$ (this
can be done by adjusting $c_0$). In this way one finds $a_1 \approx 0.1289$ and $a_{-1} \approx 0.1263$.
The associated $b_l$'s are $b_1 \approx 0.1234$ and $b_{-1} \approx 0.1262$. Using these
one find that the mass ratios in \eqref{mass4} are satisfied. 
Next choosing $\eta \approx 0.223$ so that $a_0 \approx 0.223$ and $b_0 \approx 1.397$
one finds that the mixings in \eqref{ckmratio} and \eqref{ckmratioa} are satisfied.

Other values of $a$, $b$, $c$, $D$ and $x$ in this general range worked as well. However
in general the various relationships worked best when $b$ was close to $4$. 

\section{Summary and Conclusions}

In this paper we studied the field equations of fermions in the 
background of the non-singular, 6D brane solution of \cite{Mi} \cite{GoSi} .
By allowing the exponent, $b$, in the 4D scale function, $\phi (r)$, to
take values $b >2$ we found that we could get multiple zero mass
modes which were identified with different fermion generations.
In particular for $4< b \le 8$ we obtained three zero mass modes corresponding
to different $l$ eigenvalues: $l=-1, 0, 1$. The charge $l$ played the role
of the family number. When one fixes the value
of $b=2$ as in \cite{GoSi} one has only one zero mode \cite{Ma}.
For $b>2$ the 2D scale function, $\lambda$ has 
a zero both at $r=0$ and $r= \infty$. However the scalar invariants such as the
Ricci scalar are well behaved and non-singular over the entire range of $r$,
indicating these points are coordinate rather than physical singularities.

The masses and mixings between the different generations
was given by a common mechanism -- the introduction of a
scalar field with a Yukawa coupling to the fermions. 

An interesting extension of the above scenario is to see if
the scalar field could play a dual role: (i) as the mechanism for
generating the masses and mixings and (ii) as the matter source 
for forming the brane. In \cite{Br} \cite{Br1} it was shown that 
a scalar field could be used as a source to construct a thick brane. Thus
it might be possible to replace the phenomenological matter sources, 
$F(r)$ and $K(r)$, by a scalar field source. Such a scenario would be more
economical since the scalar would serve the dual role of forming the brane 
and giving masses and mixings to the fermions.

\begin{flushleft}
{\bf Acknowledgments} DS acknowledges the CSU Fresno College of
Science and Mathematics for a sabbatical leave during the period
when part of this work was completed. DS also thanks Prof. Vitaly Melnikov 
for the invitation to work at VNIIMS and the People's Friendship
University of Russia where part of this work was completed.
\end{flushleft}


\end{document}